\documentclass[prd,amsfonts,amsmath,twocolumn,superscriptaddress,nofootinbib]{revtex4}

\usepackage{color}
\usepackage{graphicx} 
\usepackage{epstopdf}%
\usepackage{placeins}
\usepackage{amssymb}
\usepackage{cleveref}
\usepackage{mathpazo}      

\newcommand {\be} {\begin{equation}} 
\newcommand {\ba}{\begin{Eqarray}} 
\newcommand {\ee} {\end{equation}} 
\newcommand{\ea} {\end{Eqarray}}

\renewcommand{\epsilon}{\varepsilon}

\begin{document}

\title{Radiative Capture of Cold Neutrons by Protons and \\ Deuteron Photodisintegration with Twisted Beams}

\author{Andrei Afanasev}

\affiliation{Department of Physics,
The George Washington University, Washington, DC 20052, USA}

\author{Valery G. Serbo}
\affiliation{Novosibirsk State University, RUS-630090, Novosibirsk, Russia}
\affiliation{Sobolev Institute of Mathematics, RUS-630090, Novosibirsk, Russia}

\author{Maria Solyanik$^1$}

\date{\today}

\begin{abstract}

We consider two basic nuclear reactions: Radiative capture of neutrons by protons, $n+p\to \gamma+~d$ and its time-reversed counterpart, 
photodisintegration of the deuteron, $\gamma +d\to n+p$. In both of these cases we assume that the incoming beam of neutrons or photons is ``twisted" by having an azimuthal phase dependence, {\it i.e.}, it carries an additional angular momentum along its direction of propagation. Taking a low-energy limit of these reactions, we derive relations between corresponding transition amplitudes and cross sections with plane-wave beams and twisted beams. Implications for experiments with twisted cold neutrons and photon beams are discussed.
\end{abstract}

\maketitle

\section{Introduction}	\label{sec:intro}
Let us consider two types of nuclear reactions: radiative capture of cold neutrons and photodisintegration of the deuteron.
Both of these reactions with energies $E \lesssim 1 ~$MeV by protons is of high importance for Big Bang nucleosynthesis and stellar evolution.  

Radiative capture of neutrons was considered in great detail in a number of experimental and theoretical papers, {\it c.f.} the following review article \cite{rupak2000precision}. We are particularly interested in reactions with low neutron energies $E \sim 10$ meV, that can be tested in experiments with cold neutrons. The total reaction cross-section is $\sigma \sim 300$ mb, and mainly comes from $M1$ isovector contribution dominating in $S$ wave capture.
{\it Twisted cold neutrons} in this energy range have been recently generated in the lab \cite{clark2015controlling} (with temperature $130$K and wavelength $\lambda = 0.27$ nm). The neutron beams with Orbital Angular Momentum (OAM) projection $\hbar m$ on the direction of beam propagation were obtained, where $m=1, 2, 4$. It is reasonable to expect these neutrons to be applied to $np \rightarrow d\gamma$ reaction studies. Thereby, this reaction can be utilized in diagnostics of neutron beams' topology. 

Feasibility of obtaining {\it twisted photon} beams with energies in the MeV range sufficient to induce photo-nuclear reactions was theoretically demonstrated in the literature \cite{jentschura11gen,jentschura11com}, from electrons in spiral motion related to astro-physics conditions \cite{Katoh2017}. Simulations for specific experimental conditions for high-energy twisted-photon generation was provided in Ref.\cite{Petrillo2016} for Compton backscattering and in Ref.\cite{Liu2016} for plasma-wakefield acceleration. We may also expect generation of twisted photons in a non-linear regime of Compton scattering with circularly-polarized initial photons \cite{taira2016gamma}. 

In view of the technological progress in generation of twisted beams, we need to evaluate potential importance of such beams for nuclear studies.
Here we consider theoretically the simplest nuclear reactions caused by twisted neutron or photon beams.
In particular, in this paper we discuss distinctive features of the twisted neutron beams' capture by macroscopic and mesoscopic proton targets, which can be experimentally observed.  Its time-reversed analogue, deuteron photodisintegration is considered for the case of low energies, when the S-state dominates in the partial-wave expansion. 

The paper is organized as follows. Section \ref{sec:2} introduced the formalism of twisted-neutron beams, Section \ref{sec:3} discusses the reaction of radiative capture with twisted neutrons for different types of targets, Section \ref{sec:4}  presents comparison between nuclear photoabsorption of the twisted and plane-wave photons, and conclusions are presented in Section \ref{sec:summary}.

\section{Incoming Twisted Neutron}	\label{sec:2}

We use the formalism of Quantum Mechanics to describe the corresponding baryonic states, neglecting the neutron spin degree of freedom. 
Consider a stationary state of free twisted neutron of energy $E$ and mass $m_n$ in the non-relativistic regime. We take $\hbar = 1$ for simplicity. The wave function of such a state in the cylindrical coordinate system $\rho, \phi_r, z$ can be written as an eigenfunction of three commuting operators: energy $\hat{H} = \frac{\hat{\pmb{p}}^2}{2m_n}$, momentum $\hat{p}_z = -i\frac{\partial}{\partial z}$ and $\hat{L}_z = -i\frac{\partial}{\partial \phi_r}$ projected on $z$-axis. Consequently, this wave function satisfies the following equations
\begin{equation}
\hat{H} \psi (\pmb{r}) = E \psi (\pmb{r}),\;\;\; \hat{p}_z \psi (\pmb{r}) = p_z \psi (\pmb{r}),\;\;\;\hat{L}_z \psi (\pmb{r}) = m \psi(\pmb{r}),
\end{equation}
where $m$ is an integer. Both longitudinal momentum and the norm of transverse momentum with respect to the $z$-direction
\begin{equation}
\varkappa = \sqrt{2m_nE - p_z^2}
\end{equation}
are well defined for this state. Then the explicit functional from is
\begin{equation}
\psi_{\varkappa m p_z}(\pmb{r}) = J_m (\varkappa \rho) e^{im\phi_r} e^{ip_z z},
\label{04/26/17_1}
\end{equation}
where $J_m(x)$ is the Bessel function of the first kind. The plane wave expansion of this expression in the $xy$-plane can be written as following
\begin{equation}
\psi_{\varkappa mp_z}(\pmb{r}) = e^{ip_z z} \int a_{\varkappa m}(\pmb{p}_{\perp}) e^{i\pmb{p}_{\perp} \pmb{\rho}}\frac{d^2 p_{\perp}}{(2\pi)^2}
\end{equation}
where $a_{\varkappa m}(\pmb{p}_{\perp})$ is the corresponding Fourier amplitude
\begin{equation}
a_{\varkappa m}(\pmb{p}_{\perp}) = i^{-m} \frac{2\pi}{p_{\perp}} \delta(p_{\perp} - \varkappa) e^{im\phi_p}
\label{04/28/2017_1}
\end{equation}
Here $\phi_{p}$ is the azimuthal angle of vector $\pmb{p} = (\pmb{p}_{\perp}, p_z)$ and $p_{\perp} = |\pmb{p}_{\perp}|$. The Fourier amplitude is normalized as
\begin{equation}
\int a_{\varkappa m}^*(\pmb{p}_{\perp}) a_{\varkappa' m'}(\pmb{p}_{\perp})\frac{d^2 p_{\perp}}{(2\pi)^2} = \frac{2\pi}{\varkappa} \delta(\varkappa - \varkappa') \delta_{mm'}
\end{equation}
from which the neutron state normalization condition follows:
\begin{equation}
\begin{split}
\int \psi_{\varkappa m p_z}^*(\pmb{r}) \psi_{\varkappa' m' p'_z}(\pmb{r}) d^3 r= \;\;\;\;\;\;\;\;\;\;\;\;\;\;\;\;\;\;\;\;\;\;\;\;\;\;\;\;\;\;\; \\ \int a_{\varkappa m}^*(\pmb{p}_{\perp}) a_{\varkappa' m'}(\pmb{p}_{\perp}) e^{i(\pmb{p} - \pmb{p}')\pmb{r}} \frac{d^2 p_{\perp} d^2 p'_{\perp}}{(2\pi)^4} d^3 r = \\ \frac{4\pi^2}{\varkappa} \delta(\varkappa - \varkappa') \delta_{mm'}\delta(p_z - p'_z).
\end{split}
\end{equation}
Hence, twisted neutron can be seen as a non-diverging packet of plane waves, localized on a conical surface with the opening angle $\theta_p = \arctan(\varkappa/p_z)$ and symmetry axis aligned with the $z$-direction.

In the following, we consider the limiting case of Eq. \eqref{04/26/17_1} for fixed energy $E$ and $\theta_p \rightarrow 0$ (in this case $\varkappa \rightarrow 0$, $p_z \rightarrow \sqrt{2m_n E}$ and $J_m(\varkappa \rho) \rightarrow \delta_{m0}$):
\begin{equation}
\psi_{\varkappa mp_z} (\pmb{r})|_{\theta_p \rightarrow 0} = \delta_{m0} e^{ipz}.
\end{equation}
In other words, in this limit one recovers a standard expression for the plane wave propagating along the $z$-axis.

The density of a twisted neutron state propagating in the $z$-direction is
\begin{equation}
\varrho^{(m)}(\rho) = |\psi_{\varkappa m p_z}(\pmb{r})|^2 = J^2_m(\varkappa \rho),
\label{04/28/2017_4}
\end{equation}
see Fig. 1. This is the positive-definite function of coordinates in the transverse plane which depends on the quantum number $m$. It is worth it to mention that
\begin{equation}
\varrho^{(m)} (\rho)|_{\theta_p \rightarrow 0} = \delta_{m0}.
\end{equation}
Hence, for $m=0$ one recovers standard expression for the density in this limit. This density is non-zero at the origin for $m=0$,
\begin{equation}
\varrho^{(0)}(\rho)|_{\rho \rightarrow 0} = 1.
\end{equation}

\begin{figure}
\centering
\includegraphics[scale=0.8]{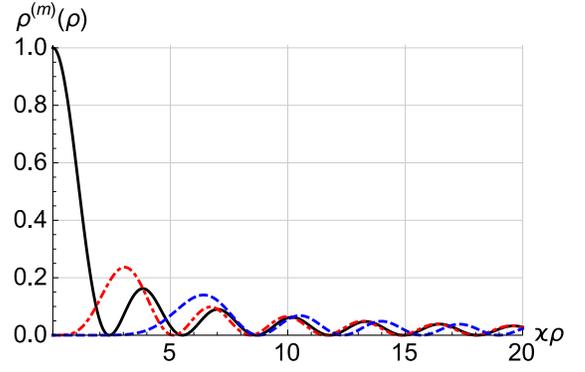}
\caption{Function $\varrho^{(m)}(\rho)$, Eq. \eqref{04/28/2017_4}, as a function of $\varkappa \rho$ for $m=0$ (black solid line), $m=2$ (red dot-dashed line), $m=5$ (blue dashed line).}
\label{fig:25/07/17_1}
\end{figure}
\FloatBarrier
Let $M^{\text{(pl)}}(\pmb{p}, \pmb{k})$ be the matrix element of plane wave neutron radiative recombination 
\begin{equation}
n(\pmb{p})+ p \rightarrow d+\gamma(\pmb{k}).
\end{equation}
As the $S$-wave contributes the most into the cross section of this process, the corresponding matrix element is independent of the direction of the incoming neutron momentum $\pmb{p}$. When considering radioactive recombination, we need to account for the fact that the twisted neutron has a symmetry axis aligned with the $z$-direction, while a proton is located at a certain distance $\pmb{b} = (b_x, b_y, 0)$ from this axis called impact parameter. Hence, we can express this matrix element as a trace of a plane-wave matrix element $M^{\text{(pl)}}(\pmb{p}, \pmb{k})$ and the Fourier amplitude $a_{\varkappa m}(\pmb{p}_{\perp})$ of the incoming electron with the impact parameter $b$
\begin{equation}
\begin{split}
M^{(m)}(\varkappa, p_z, \pmb{k}, \pmb{b}) = \;\;\;\;\;\;\;\;\;\;\;\;\;\;\;\;\;\;\;\;\;\;\;\;\;\;\;\;\;\;\;\;\;\;\;\;\;\;\;\;\;\;\;\;\;\;\;\; \\e^{-i\pmb{k}_{\perp} \pmb{b}} \int M^{\text{(pl)}}(\pmb{p}, \pmb{k}) e^{i\pmb{p}_{\perp} \pmb{b}} a_{\varkappa m}(\pmb{p}_{\perp}) \frac{d^2 p_{\perp}}{(2\pi)^2} = \\e^{im\phi_b - i\pmb{k}_{\perp} \pmb{b}}M^{\text{(pl)}}(\pmb{p}, \pmb{k})J_m(\varkappa b).
\end{split}
\end{equation}
This analysis can be applied to Bessel-Gaussian diffracting mode, where the Bessel-shaped disturbance in transverse plane is suppressed by a Gaussian factor. A conventional wave-function for BG mode in focus plane ($z=0$) can be written as \cite{doster2016laguerre}
\begin{equation}
\psi_{\varkappa m p_z}^{\text{(BG)}}(\pmb{r}) = N J_m (\varkappa \rho) \exp \Big(-\frac{\rho^2}{\text{w}_0^2} \Big) e^{im \phi_{\rho}}
\label{04/28/2017_2}
\end{equation}
where $N$ is a constant factor. In this case the corresponding amplitude acquires this additional phase factor
\begin{equation}
M^{\text{(BG)}}(\varkappa, p_z, \pmb{k}, \pmb{b}) = e^{im\phi_b - i\pmb{k}_{\perp} \pmb{b}}J_m(\varkappa b) e^{-\frac{b^2}{\text{w}_0^2}} M^{\text{(pl)}}(\pmb{p}, \pmb{k})
\end{equation}
where $\text{w}_0$ is the Gaussian waist of the BG mode.

It follows that the differential probability of the recombination process per unit time for the case of a twisted neutron $d\dot{W}^{(m)}(\pmb{b})/d \Omega_k$ and for a plane wave $d\dot{W}^{\text{(pl)}}(\pmb{b})/d \Omega_k$ are related by the following expression
\begin{equation}
\frac{d\dot{W}^{(m)}(\pmb{b})}{d \Omega_k} = \frac{d\dot{W}^{\text{(pl)}}(\pmb{b})}{d \Omega_k} J^2_m (\varkappa b)
\label{04/26/2017_1}
\end{equation}

\section{$np \rightarrow d\gamma$ reaction with cold twisted neutrons}	\label{sec:3}

\subsection{Macroscopic target} \label{subsec:3.1}
To begin with, we consider the simplest case of a macroscopic proton target with a uniform random distribution of scattering centers in the plane, normal to the direction of the beam propagation. We also neglect secondary collisions. In this setup, the differential cross section averaged over the proton impact parameter in the area with radius $R$ is the convenient observable to consider. Following paper \cite{serbo2015scattering}, we obtain a simple relation for the averaged cross section expressed as a function of the plane wave cross section:
\begin{equation}
\frac{d \bar{\sigma}}{d \Omega_k} = \frac{1}{\cos \theta_p} \int_0^{2\pi} \frac{d \sigma^{\text{(pl)}}}{d \Omega_k} \frac{d\phi_p}{2\pi},
\end{equation}
where $\theta_p$ is the conical angle of the incoming twisted neutron. The standard scattering cross section is independent of $\phi_p$ in our case, which leads to
\begin{equation}
\frac{d\bar{\sigma}}{d\Omega_k} = \frac{1}{\cos \theta_p} \frac{d \sigma^{\text{(pl)}}}{d\Omega_k}, \;\;\; \bar{\sigma} = \frac{\sigma^{\text{(pl)}}}{\cos \theta_p}.
\end{equation}
For the limit $\theta_p \rightarrow 0$, the averaged cross section is the same as the standard plane-wave one.

\subsection{Mesoscopic target} \label{subsec:3.2} \label{sec.IIIb}

In this case, we consider a proton target of finite size, uniformly and symmetrically distributed with respect to its center. The spatial distribution of protons inside this target is considered to be classical and can be described the by density function $n(\pmb{b} - \pmb{b}_t)$. As before, the impact parameter $\pmb{b}$ is defined as the distance from the neutron beam axis to a proton in the target, and vector $\pmb{b}_t$ corresponds to the distance from the beam axis to the center of the target. In this case, differential probability integrated over the impact parameters of the finite-size target is the common characteristic of the process, Eq. \eqref{04/27/2017_1}.
\begin{equation}
\int \frac{d \dot{W}^{(m)}(\pmb{b})}{d \Omega_k} n(\pmb{b} - \pmb{b}_t)d^2 b
\label{04/27/2017_1}
\end{equation}
However, an even more distinct observable is this probability divided by the incoming neutron flux $J_z^{(mesos)}$. Taking into consideration properties of proton density, it is reasonable to define twisted neutron flux density along the $z$ axis as follows
\begin{equation}
j_z^{(\text{tw})}(\rho) = \frac{p_z}{m_n} \varrho^{(0)}(\rho) = v_z J_0^2(\varkappa \rho),
\end{equation}
which behaves nicely in the plane-wave limit
\begin{equation}
j_z^{(\text{tw})}(\rho)|_{\theta_p \rightarrow 0} = \frac{p}{m_n} = v,
\end{equation}
\begin{figure}
\centering
\includegraphics[scale=0.8]{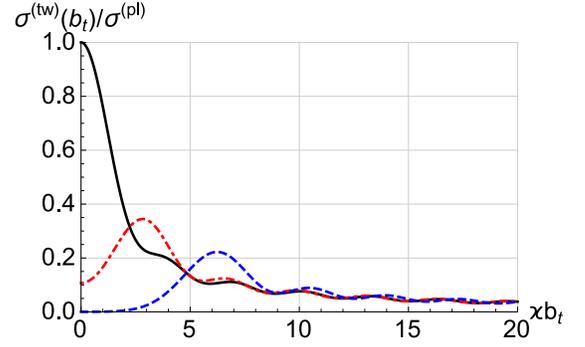}
\caption{The cross section ratio $\sigma^{\text{(tw)}}(b_t)/\sigma^{\text{(pl)}}$, Eq. \eqref{04/28/2017_3}, as a function of the impact parameter $b_t$ for $m=0$ (black solid line), $m=2$ (red dot-dashed line), $m=5$ (blue dashed line).}
\label{fig:25/07/17_2}
\end{figure}
\FloatBarrier
being non-zero at the origin. As the result we define the flux
\begin{equation}
J_z^{(mesos)} = \int j_z^{(\text{tw})} (\rho) n(\pmb{\rho}) d^2\rho
\label{09/12/2017_1}
\end{equation}
and mesoscopic cross section
\begin{equation}
\frac{d \sigma^{(mesos)}(\pmb{b}_t)}{d \Omega_k} = \frac{1}{J_z^{(mesos)}} \int \frac{d\dot{W}^{(m)}(\pmb{b})}{d\Omega_k} n(\pmb{b} - \pmb{b}_t) d^2b
\end{equation}
Taking into consideration Eq.\eqref{04/26/2017_1} we get
\begin{equation}
\frac{d\sigma^{(mesos)}(\pmb{b}_t)}{d\Omega_k} = \frac{R}{\cos \theta_p} \frac{d\sigma^{\text{(pl)}}}{d \Omega_k},\;\;\; R=\frac{\int J^2_m (\varkappa b) n(\pmb{b} - \pmb{b}_t)d^2 b}{\int J^2_0 (\varkappa b) n(\pmb{b}) d^2 b}.
\end{equation}
In what follows, we use a Gaussian distribution which is a standard choice for ion traps
\begin{equation}
n(\pmb{b} - \pmb{b}_t) = \frac{1}{2\pi w^2} e^{-\frac{(\pmb{b}-\pmb{b}_t)^2}{2w^2}}
\label{12/09/17_10}
\end{equation}
where $w$ is the dispersion characterized by the size of the proton target. In this case the integration by parts over $b$ can partially be done analytically
\begin{equation}
\begin{split}
R=\frac{1}{I_0(\varkappa^2 w^2) e^{-\varkappa^2 w^2}} \;\;\;\;\;\;\;\;\;\;\;\;\;\;\;\;\;\;\;\;\;\;\;\;\;\;\;\;\;\;\;\; \\ \int_0^{\infty} J_m^2(\varkappa b) I_0(bb_t/w^2) e^{-(b^2+b_t^2)/(2w^2)}\frac{bdb}{w^2}.
\label{04/28/2017_3}
\end{split}
\end{equation}
Total scattering cross sections are related as follows
\begin{equation}
\frac{\sigma^{(\text{tw})}}{\sigma^{\text{(pl)}}} = \frac{R}{\cos \theta_p}.
\end{equation}
In Fig. \ref{fig:25/07/17_2} this relationship is plotted as a function of displacement $b_t$ of the Gaussian target from the beam axis for different values of the quantum number $m$. The target size is taken to be $w = 1/\varkappa$ and the conical angle - $\theta_p = 1^o$. When comparing this plot to Fig. \ref{fig:25/07/17_1}, one can clearly see, that $\sigma^{(\text{tw})}(b_t)$ is very sensitive to the density distribution of the incoming twisted neutrons. Hence, this reaction may appear to be very useful in diagnostics of the twisted neutron beams.

\section{Nuclear photo-absorption} \label{sec:4}

\subsection{Deuteron Photodisintegration with Plane-Wave Photons}

In this section we will mostly follow the description of the nuclear electromagnetic interactions in \cite{deshalit1974theoretical} and \cite{walecka2004theoretical}. As is well known, the space-time solution of the EM wave equation in a Cartesian coordinate system delivers a plane-wave vector potential in the following form:
\begin{equation}
\pmb{A}^{(\text{pl})}_{\pmb{k} \lambda} (\bold{r}, t) = \boldsymbol{\epsilon}_{\pmb{k} \lambda} e^{-i(\omega t - \pmb{k} \cdot \pmb{r})}
\label{04/12/17/4}
\end{equation}
which describes the propagation of a photon with frequency $\omega$ and wave-vector $\pmb{k}$ carrying polarization $\boldsymbol{\epsilon}_{\pmb{k} \lambda}$, such that $\lambda = \pm1$ is the helicity index. Starting from here we will use the natural units: $\hbar = c = 1$. Applying the field quantization conditions we obtain the vector potential operator:
\begin{equation}
\begin{gathered}
\hat{\pmb{A}}^{\text{(pl)}}_{\pmb{k} \lambda}(\pmb{r}, t) = \frac{1}{2\pi^2} \sum_{\lambda = \pm 1} \\ \int \frac{d\pmb{k}}{\sqrt{2\omega_k}} (\boldsymbol{\epsilon}_{\pmb{k} \lambda}^* \hat{a}^{\dagger}_{\lambda} (\pmb{k}) e^{-i(\pmb{k} \cdot \pmb{r} - \omega t)}+ c.c)
\end{gathered}
\label{04/12/17/5}
\end{equation}
where $\hat{a}_{\lambda} (\pmb{k})$ and $\hat{a}^{\dagger}_{\lambda} (\pmb{k})$ are the photon annihilation and creation operators obeying the commutation relation
\begin{equation}
[\hat{a}_{\lambda} (\pmb{k}), \hat{a}^{\dagger}_{\lambda'} (\pmb{k'})] = \frac{\delta(\pmb{k} - \pmb{k}')}{kk'}\delta_{\lambda \lambda'}
\label{04/12/17/2}
\end{equation}
so that the corresponding polarization states are given as
\begin{equation}
\boldsymbol{\epsilon}_{\pmb{k} \lambda} = -\frac{\lambda}{\sqrt{2}}(\bold{e}_x + i \lambda \bold{e}_y)
\end{equation}
The choice of the polarization basis is purely conventional and has been made this way to simplify the formalism.

The interaction of the nucleus with the electromagnetic field up to leading order in charge is described by the Hamiltonian:
\begin{equation}
\hat{H}_{\text{int}} = - \int d\pmb{r}\; \hat{\pmb{j}} (\pmb{r}) \cdot \hat{\pmb{A}}^{\text{(pl)}}_{\pmb{k} \lambda}(\pmb{r}, t)
\label{04/12/17/1}
\end{equation}
where $\hat{\pmb{j}} (\pmb{r})$ is the nuclear EM current
\begin{equation}
\hat{\pmb{j}}(\pmb{r}) = \hat{\pmb{j}}_c + \hat{\pmb{j}}_m
\end{equation}
consisting of two contributions. The convection current
\begin{equation}
\hat{\pmb{j}}_c = \sum_i \frac{1}{4} e(1+\tau_3 (i))(\hat{\pmb{v}}_i \delta(\pmb{r} - \pmb{r}_i) + \delta(\pmb{r} - \pmb{r}_i)\hat{\pmb{v}}_i)
\end{equation}
comes from the relative motion of protons, while the magnetization current comes from the spin orientation of the nucleons in the nucleus
\begin{equation}
\hat{\pmb{j}}_m = \frac{e}{4m_p} \sum_i \frac{1}{2} [(g_p + g_n) + \tau_3(i) (g_p - g_n)] \cdot \nabla \times [\sigma_i \delta(\pmb{r} - \pmb{r}_i)]
\end{equation}
where the index $i$ counts the nucleons; $\sigma_i$ is the set of spin Pauli matrices; $\pmb{\tau}$ is the isospin operator in SU(2); $e$ and $m_p$ are proton charge and mass; and $g_p$, and $g_n$ are proton and neutron spin g-factors respectively.

Making use of Fermi Golden Rule, we can calculate the nuclear photo-absorption probability
\begin{equation}
dW(N_i \omega \rightarrow N_f) = 2\pi k^2 |M_{fi}|^2 d \Omega
\end{equation}
where $d\Omega$ is a solid angle of one of the final nucleon states, $d\Omega_n$ or $d\Omega_p$. The key element is the photo-absorption amplitude
\begin{equation}
M^{\text{(pl)}}_{jm} = \langle J_f M_f | H_{\text{int}} | J_i M_i; \pmb{k} \lambda \rangle
\label{05/04/2017_1}
\end{equation}
with $j$ and $m$ indicating the quantum numbers of the total angular momentum of the system and its projection on the quantization axis correspondingly. If one considers the interaction Hamiltonian \eqref{04/12/17/1} and applies it to the amplitude the resulting expression
\begin{equation}
M^{\text{(pl)}}_{jm} = -\sqrt{\frac{1}{4\pi^2 \omega_k}} \langle J_f M_f | \int d\pmb{r} \; \hat{\pmb{j}}(\pmb{r}) \cdot \boldsymbol{\epsilon}_{\pmb{k} \lambda} e^{-i(\pmb{k} \cdot \pmb{r} - \omega t) } | J_i M_i \rangle 
\label{04/12/17/3}
\end{equation}
where the commutation relation \eqref{04/12/17/2} has been used to annihilate the incoming photon state. The incoming plane wave photon has a well-defined quantization axis, which makes it only logical to reduce the complexity of the problem by making a choice of the preferred direction such that $\pmb{k}$ is parallel to the axis $z$. This allows us to make the following simplifications \cite{varshalovich1988quantum}:
\begin{equation}
e^{i\pmb{k} \cdot \pmb{r}} \rightarrow e^{ikz} = \sum_{\ell} i^{\ell} \sqrt{4\pi(2\ell + 1)} j_{\ell} (kr) Y_{\ell 0}(\Omega_k)
\end{equation}
Rewriting the product of the polarization state with the exponential factor and making use of the vector spherical harmonics $\bold{Y}_{j\ell 1}^{m}(\Omega_k)$ we get
\begin{equation}
\begin{gathered}
\boldsymbol{\epsilon}_{\bold{k}\lambda} e^{i\pmb{k} \cdot \pmb{r}} \rightarrow \boldsymbol{\epsilon}_{\bold{k}\lambda} e^{ikz} = \sum_{\ell} i^{\ell} \sqrt{4\pi (2\ell + 1)} \\ j_{\ell}(kr) \langle \ell 0 1 \lambda | j \ell \rangle \bold{Y}_{j\ell 1}^{\lambda}(\Omega_k)
\end{gathered}
\end{equation}
where the symmetry properties of the Glebsch-Gordan coefficients have been used to impose the quantization condition $m = \lambda$. That means that the corresponding absorption amplitude \eqref{04/12/17/3} is independent of the magnetic quantum number $m$, but depends on its component $\ell$ instead:
\begin{equation}
\begin{gathered}
M_{j m}^{\text{(pl)}} (0)  \rightarrow M_{j \lambda}^{\text{(pl)}} (0)= -\sqrt{\frac{1}{4\pi^2 \omega_k}} \sum_{\ell} i^{\ell} \sqrt{4\pi(2\ell+1)} \\ \langle \ell 0 1 \lambda | j \lambda \rangle e^{-i\omega_k t} \int d\pmb{r} \; \langle J_f M_f | j(kr) (\bold{Y}_{j\ell 1}^{\lambda}(\Omega_k) \cdot \hat{\pmb{j}} (\pmb{r})) | J_i M_i \rangle
\end{gathered}
\end{equation}
where zero in the argument $M_{j \lambda}^{\text{(pl)}} (0)$ indicates that the incoming photon is directed in solid angle $\theta_k = \phi_k=0$ with respect to $z$. If we now apply Wigner - Eckart theorem to this matrix element, we get
\begin{equation}
\begin{gathered}
\langle J_f M_f | \mathcal{M}_{j\lambda} (0) | J_i M_i \rangle = \langle J_i M_i\; j \lambda | J_f M_f \rangle \langle J_f || \mathcal{M}_{j \lambda} (0) || J_i \rangle
\end{gathered}
\end{equation}
This results into the set of selection rules:
\begin{equation}
\begin{gathered}
|J_i - J_f| \leq |j| \leq J_i + J_f;\\
M_f - M_i = \lambda
\end{gathered}
\label{04/27/2017_2}
\end{equation}
which tells us that the full projection of the incident photon spin in the direction of propagation gets absorbed by the nucleus.

As our ultimate goal is to apply this formalism for the case of the twisted light, we need one more modification related to the topological structure of the beams with OAM. Any twisted beam can be Fourier transformed, hence, expressed as the ensemble of plane waves with the wave-vectors directed under certain solid angle $\Omega_k$. This angle can be decomposed into opening angle $\theta_k$ and azimuthal angle $\phi_k$ to the direction of quantization axis $z$. However, instead of rotating the photon state, which would result into major mathematical complications, we will rotate the nuclear states by the same angles \cite{varshalovich1988quantum}, such that the initial and final states of the nucleus are expressed
\begin{equation}
\begin{gathered}
|J_i M_i \rangle = \psi_{J_i M_i} (\pmb{r}') = \sum_{M'_i} D_{M_i M'_i}^{J_i*} (\Omega_k) \psi_{J_i M_i} (\pmb{r});\\
|J_f M_f \rangle = \psi_{J_f M_f} (\pmb{r}') = \sum_{M'_f} D_{M_f M'_f}^{J_f*} (\Omega_k) \psi_{J_f M_f} (\pmb{r}).
\end{gathered}
\end{equation}
This spatial transformation leads to the following expression for the amplitude
\begin{equation}
M_{j \lambda}^{\text{(pl)}} (\Omega_k) = -\sum_{M'_f M'_i} D_{M_i M'_i}^{J_i*} (\Omega_k)D_{M_f M'_f}^{J_f}(\Omega_k) M_{j \lambda}^{\text{(pl)}} (0).
\end{equation}
It is worth mentioning that the isospin selection rules are not discussed here since they are not modified by the topological structure of the beam. We also assumed that parity is conserved, leaving aside the weak parity violation effects.
\\
\subsection{Deuteron Photodisintegration with twisted photons} \label{sec:5}

In this subsection we will consider a deuteron interacting with a twisted beam of photons with well-defined TAM $m_{\gamma}$, in particular, BB and BG. In scalar formulation, Bessel mode is the exact solution of the wave equation in cylindrical coordinates (non-paraxial regime) \cite{durnin1987diffraction}. BG mode \cite{gori1987bessel} belongs to the Gaussian family solutions of the paraxial wave equation (paraxial regime $|\varkappa| \ll |\pmb{k}|$).

Following the Bessel mode formalism, laid out in Sec. \ref{sec:2}, one can express the twisted photon state as a 2D Fourier transform
\begin{equation}
|\varkappa m_{\gamma} k_z \lambda \rangle = e^{-i(\omega t - k_z z)} \int a_{\varkappa m_{\gamma}} (\pmb{k}_{\perp}) e^{i \pmb{k}_{\perp} \cdot \pmb{\rho}} \frac{d^2 k_{\perp}}{(2\pi)^2}
\end{equation}
with the Fourier kernel \eqref{04/28/2017_1} in case of BB. The vector potential of this state can be written as
\begin{equation}
\pmb{A}^{\text{(BB)}} = e^{-i(\omega t - k_z z)}\int a_{\varkappa m_{\gamma}}(\pmb{k}_{\perp}) \boldsymbol{\epsilon}_{\pmb{k}\lambda}e^{i \pmb{k}_{\perp} \cdot \pmb{\rho}} \frac{d^2 k_{\perp}}{(2\pi)^2}
\end{equation}
which, making use of \eqref{04/12/17/4} and \eqref{04/12/17/5}, can be recast into
\begin{equation}
\pmb{A}^{\text{(BB)}} = \int \frac{d^2 k_{\perp}}{(2\pi)^2} a_{\varkappa m_{\gamma}}(\pmb{k}_{\perp}) \pmb{A}^{\text{(pl)}}_{\pmb{k} \lambda}(\pmb{r}, t)
\end{equation}
Here the superscript BB refers to the Bessel Beam of incident photons. If one takes into consideration the impact parameter, we get
\begin{equation}
\pmb{A}^{\text{(BB)}} = e^{-i(\omega t - k_z z)} \int \frac{d\phi_k}{2\pi} (-i)^{m_{\gamma}}e^{im_{\gamma} \phi_k} \boldsymbol{\epsilon}_{\pmb{k} \lambda}^{\mu} e^{i\pmb{k}_{\perp} \cdot (\pmb{\rho}-\pmb{b})}
\end{equation}
where $b=|\pmb{b}|$ is a photon impact parameter. As long as the matrix element \eqref{05/04/2017_1} is linear in electromagnetic field, this formula is valid.

Hence, the photo-absorption amplitude for the case of twisted photons, similar to the photo-absorption on a hydrogen-like atom \cite{scholz2014absorption}, becomes
\begin{equation}
\begin{gathered}
M^{\text{(BB)}}_{j\lambda} = -(-i)^{2m_{\gamma} - M_f + M_i} J_{m_{\gamma}-M_f + M_i} (\varkappa b) \\ e^{i(m_{\gamma}-M_f+M_i)\phi_b} \sum_{M'_i M'_f} d_{M_i M'_i}^{J_i} (\theta_k) d_{M_f M'_f}^{J_f} (\theta_k)M_{j \lambda}^{\text{(pl)}} (0)
\end{gathered}
\label{05/03/2017_1}
\end{equation}
where we have taken advantage Wigner d-function property $D_{MM'}^{J} (\Omega_k)= e^{-iM\phi_k} d_{MM'}^{J} (\theta_k)$. Here the factorization, which is spin-dependent, as opposed to the case of photo-absorption on hydrogen-like atom, ex. \cite{afanasev2016high}, is obtained for the case of photo-disintegration of deuterium.

\begin{figure*}
\vspace{2mm}
\centering
\includegraphics[scale=0.35]{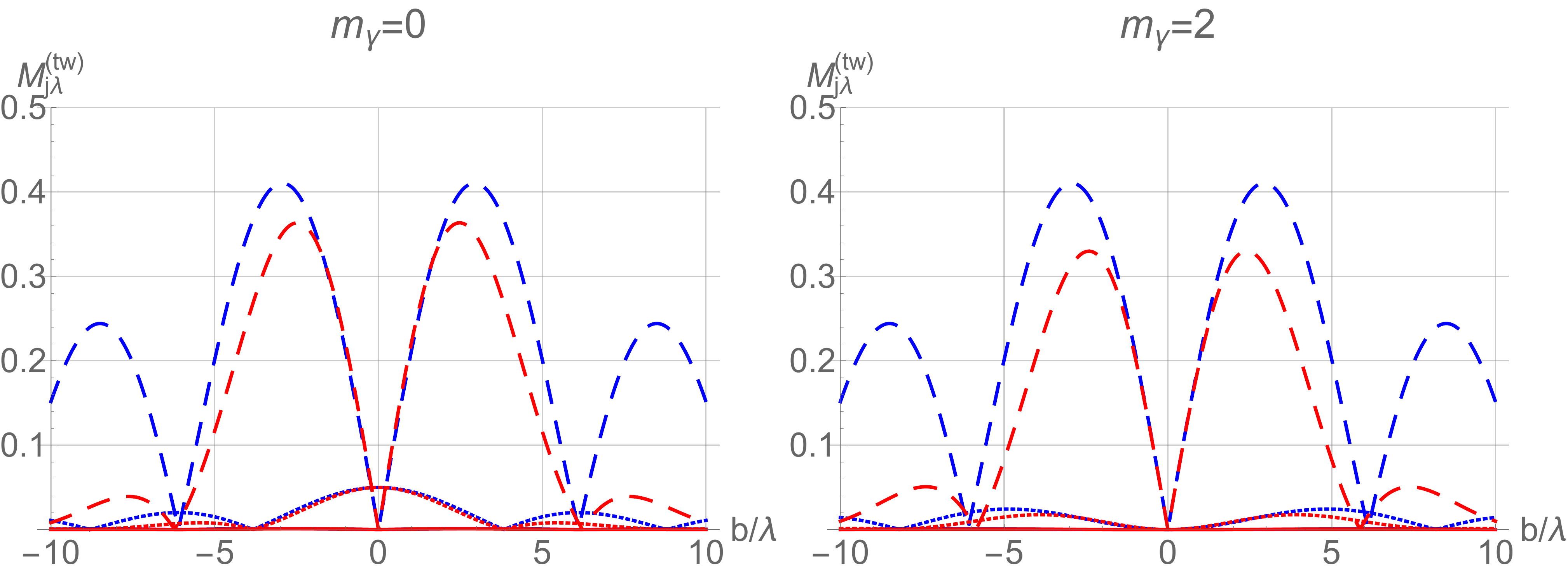}\\
\includegraphics[scale=0.35]{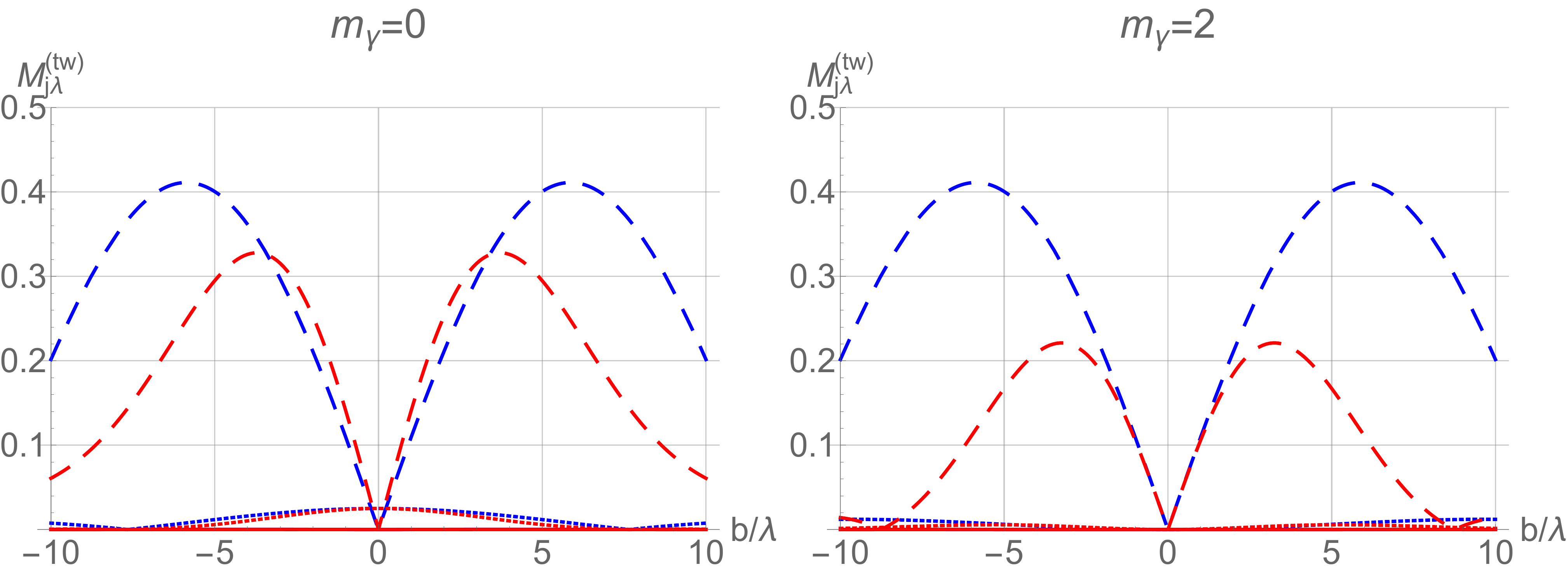}

\caption{Relative photo-absorption amplitudes $M^{\text{(tw)}}_{j\lambda}$ for $M1\to ^1S_0$ transition for BB (blue), Eq. \eqref{05/03/2017_1}; and for BG (red), Eq. \eqref{04/28/2017_6} as functions of the impact parameter scaled with the photon wavelength. The pith angles considered are $\theta_k = 0.1$~rad - top row; $\theta_k = 0.05$~rad - bottom row; and waist $\text{w}_0/\lambda = 6.0$. Dashed lines correspond to the case $M_f-M_i = 1$, dotted lines to $M_f-M_i=0$, and solid lines to $M_f-M_i=-1$.}
\label{fig:25/07/17_3}
\end{figure*}

As it is traditionally discussed elsewhere, ex. \cite{fermi1935recombination, akhiezer1950some}, the dominant contribution into the deuteron photo-disintegration comes from the transition $^3S_1 \rightarrow \; ^1S_0$ driven by M1-photons. Taking into consideration quantum selection rules \eqref{04/27/2017_2}, only one type of standard plane-wave amplitudes is contributing in this case, namely $M^{(\text{pl})}_{j\lambda} (m_i \rightarrow m_f) = M^{(\text{pl})}_{j \lambda} (-\lambda \rightarrow 0)$, which is solely due to magnetic dipole contribution: $M^{(\text{pl})}_{j \lambda} (-\lambda \rightarrow 0) = M1/\sqrt{2}$ in our normalization. As a consequence the M1 contribution in the reaction transition rate factorizes out
\begin{equation}
r(\pmb{b}) =2 |M1|^2 j^{(\gamma)}(\pmb{b})
\end{equation}
making it proportional to the flux of the incoming gamma-radiation
\begin{equation}
\begin{split}
j^{(\gamma)}(\pmb{b}) = \cos(\theta_k)\frac{\kappa \omega^2}{2\pi} \Big( J^2_{m_{\gamma}+\lambda} (\kappa b) |d_{1-1}^{1}(\theta_k)|^2 + \\+ J^2_{m_{\gamma}-\lambda} (\kappa b) |d_{-1-1}^{1}(\theta_k)|^2 + J^2_{m_{\gamma}}(\theta_k b) |d_{0-1}^{1}(\theta_k)|^2 \Big)
\label{09/12/2017_2}
\end{split}
\end{equation}

The corresponding amplitude for the twisted photon beam with Gaussian modulation (BG mode), as before \eqref{04/28/2017_2}, is expressed in \eqref{04/28/2017_6}.
\begin{equation}
\begin{gathered}
M^{\text{(BG)}}_{j\lambda} = - N \text{w}_0^2 (-i)^{M_i-M_f} e^{-\frac{\varkappa^2 \text{w}_0^2}{4}} e^{i(m_{\gamma}-M_f+M_i)\phi_b} \\ \sum_{M'_fM'_i} d_{M_f M'_f}^{J_f} (\theta_k) d_{M_i M'_i}^{J_i} (\theta_k) M_{j \lambda}^{\text{(pl)}} (0) \\  \int_0^{\infty} \frac{k_{\perp} dk_{\perp}}{2\pi} e^{-\frac{k_{\perp}^2\text{w}_0^2}{4}} I_{m_{\gamma}} \Big( \frac{\varkappa \text{w}_0^2}{2} k_{\perp} \Big) J_{m_{\gamma} - M_f+M_i} (k_{\perp} b)
\end{gathered}
\label{04/28/2017_6}
\end{equation}
where $I_m(z)$ is a modified Bessel function and the superscript (BG) stands for a Bessel-Gaussian mode, as previously in \eqref{04/28/2017_2}. The integral over transverse momentum can be taken analytically, ex. \cite{gradshteyn2014table} (6.633 1). It also can be simplified for large $\text{w}_0$ such that
\begin{equation}
\begin{gathered}
M^{\text{(BG)}}_{j\lambda} = -\frac{N}{\pi} e^{-b^2/\text{w}^2_0} (-i)^{M_i - M_f} J_{m_{\gamma} - M_f + M_i} (\kappa b) \\ e^{i(m_{\gamma} - M_f + M_i)\phi_b} \sum_{M'_f M'_i} d_{M_f M'_f}^{J_f} (\theta_k) d_{M_i M'_i}^{J_i} (\theta_k)M_{j \lambda}^{\text{(pl)}} (0)
\label{04/28/2017_7}
\end{gathered}
\end{equation}

Two distinctive coefficients, the Bessel function and the Wigner d-function, Eqs. \eqref{05/03/2017_1} and \eqref{04/28/2017_6}, are the signatures of the topological structure of the incoming photon. As it is in the case for photo-absorption on hydrogen-like atoms, ex. \cite{afanasev2013off}, these cause modification of the nuclear photo-absorption selection rules at the beam axis
\begin{equation}
M_f - M_i=m_{\gamma}
\end{equation}
as enforced by $J_{m_{\gamma} + m_i - m_f} (\varkappa b)$. The plane-wave selection rules \eqref{04/27/2017_2} are dominating on the beam periphery. This means that for the nucleus at the beam axis to absorb a photon, the entire value of TAM projection on the $z$-direction should get transferred to the nuclear degrees of freedom.

In Fig. \ref{fig:25/07/17_3} we plotted the relative photo-absorption amplitudes for the dominating transition of M1 going into $^1S_0$ finite state for the reaction $\gamma+d \rightarrow np$. We have considered initial photon states with total angular momentum $m_{\gamma} = 0, 2$ and positive helicity $\lambda = 1$. The superscript (tw) refers to a twisted incoming beam of photons (BB or BG). We get three possible transitions corresponding to three projections $M_i = -1,0,1$ of initial TAM. As one can see for the non-paraxial Bessel mode and its paraxial modification - BG mode, the amplitudes behave identically near the beam center. On the periphery $b/\lambda \gtrsim 2$, BG amplitudes get suppressed by the Gaussian factor, which fixes the unphysical infinite energy problem, present in BB. We emphasize that the details of the initial wave function of the deuterium did not affect the derived selection rules for twisted photo-absorption.

\begin{figure}
\vspace{2mm}
\centering
\includegraphics[scale=0.65]{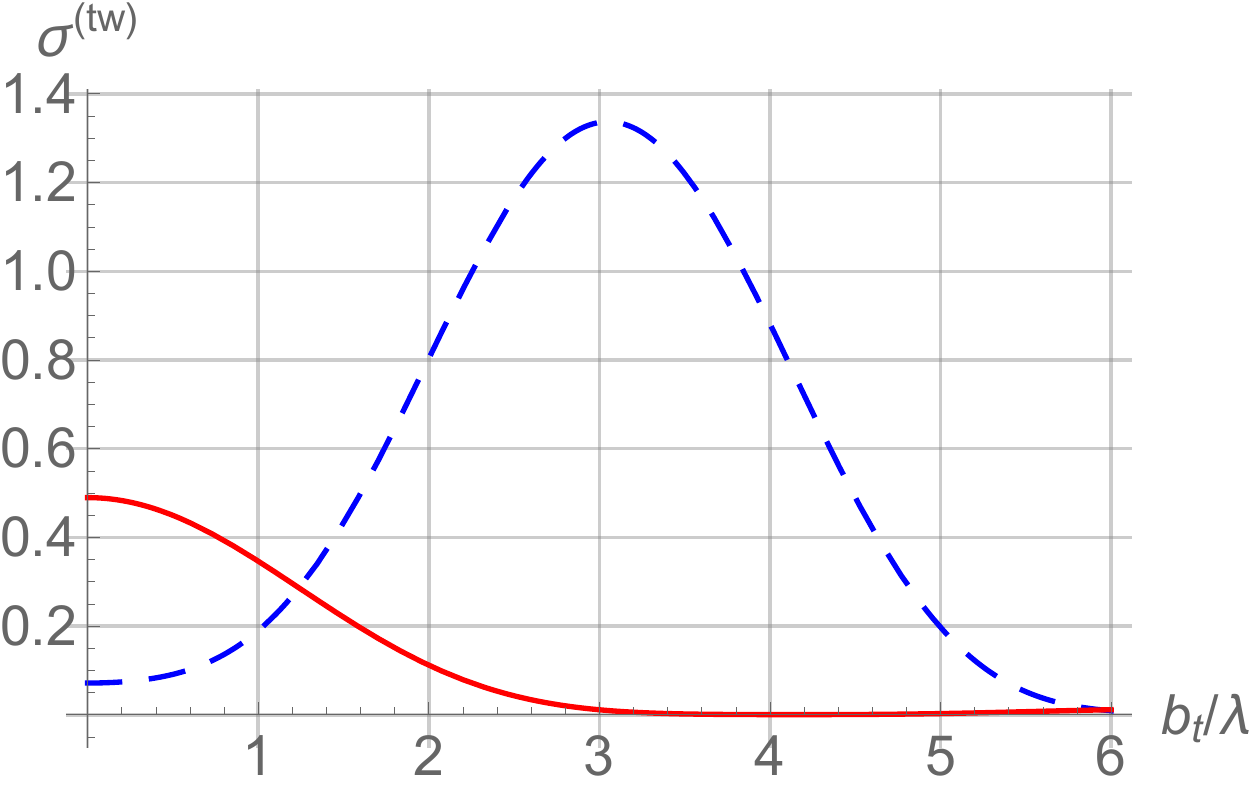}

\caption{Total cross section in arbitrary units, as a function of the position of the target center $b_t$ for the $^3S_1 \rightarrow \;^1S_0$ process, excited by twisted photon beam. Dashed-blue line corresponds to ($m_{\gamma}=0, \lambda=1$), and solid-red line to ($m_{\gamma}=0, \lambda=1$). Pitch angle is $\theta_k = 0.1$~rad.}
\label{fig:12/09/17_1}
\end{figure}

The results similar to Sec. \ref{sec:3} for mesoscopic target, with the limiting case of $w$ being much bigger than the probe's wavelength, can be worked out for the case of deuteron photodisintegration. We used Gaussian spatial distribution \eqref{12/09/17_10} to calculate the density-matrix of the excited nucleons, in analogy to \cite{peshkov2016absorption}. This leads to the equation for the mesoscopic scattering cross section in the form
\begin{equation}
\begin{gathered}
\sigma^{\text{(BB)}} (b)= \frac{2\pi \delta(\omega+E_i - E_f)}{J^{\text{(BB)}} (2J_i +1)} \sum_{m_f m_i} M_{j \lambda}^{\text{(pl)}} (\theta_k) \\ \int J_{m_{\gamma}-M_f+M_i}^2 (\kappa b) I_0\Big(\frac{bb_t}{w^2} \Big) e^{\frac{-b^2+b_t^2}{2w^2}}\frac{bdb}{w^2}
\end{gathered}
\end{equation}
for the case of Bessel photons \eqref{05/03/2017_1}, where $J^{\text{(BB)}}$ is the flux of the twisted photon beam incident on the nuclear target, which can be calculated by substitution of Eq.\eqref{09/12/2017_2} into Eq.\eqref{09/12/2017_1}. For the case of BG, the approximated from Eq.\eqref{04/28/2017_7} can be used in the similar way
\begin{equation}
\begin{gathered}
\sigma^{\text{(BG)}} (b) = \frac{2N \pi \delta(\omega+E_i - E_f)}{\pi^2J^{\text{(BG)}}(2J_i +1)} \sum_{m_f m_i} M_{j \lambda}^{\text{(pl)}} (\theta_k) \\ \int J_{m_{\gamma}-M_f+M_i}^2 (\kappa b) I_0\Big(\frac{bb_t}{w^2} \Big) e^{\frac{-3b^2+b_t^2}{2w^2}}\frac{bdb}{w^2}
\end{gathered}
\end{equation}
In analogy to the analysis in Sec. \ref{sec:3} the cross section is sensitive to the topology of the incoming photon beam, see Fig.\ref{fig:12/09/17_1}, which can be used as a diagnostic tool. However, photon wavelength for energy range of interest ($\gtrsim2.2$~MeV) is orders of magnitude smaller than a typical
mesoscopic target, in contrast to the angstrom-scale de Broglie wavelength of thermal neutron beams.

\section{Summary and Outlook} \label{sec:summary}

We analyzed theoretically the transition amplitudes and cross sections of two nuclear reactions: Radiative capture of neutrons by protons, $n+p\to \gamma+~d$ and its time-reversed analogue, 
photodisintegration of the deuteron, $\gamma +d\to n+p$. The relations between the twisted-wave and plane-wave amplitudes are derived. The results justify dedicated measurements of the nuclear reactions with the beams of twisted particles that bring additional angular momentum into the the dynamics of nuclear interactions.

\section*{Acknowledgements}

The authors are grateful to V.F. Dmitriev, C. Clark and D. Pushin for useful discussions. V.G.S. is supported by the Russian Foundation for Basic Research via
Grant No. 16-02-0644. Work of A.A. and M.S. was supported in part by Gus Weiss Endowment of George Washington University.

\bibliography{Master_b}
\bibliographystyle{ieeetr}

\end{document}